# Governing Well in the Algorithmic Age: The Foundations of Digital Statecraft

Zeynep Engin*[a,b], Tim Gordon[a], Viviana Bastidas[c], Tom Crick[d,e], Jon Crowcroft[f,g,a], Jean-Martin Denis[a,h], David J. Hand[i,a], Ed Humpherson[j], Lauren Maffeo[a,k], Jakob Mökander[l], Irene Ng[m,n], Anastasija Nikiforova[o,a], Giulio Quaggiotto[a,p], David Uriel Socol de la Osa[q,r,a], Rhonda Syler[a,s], Philip Treleaven[b,a], Stefaan Verhulst[t,u,a]

## Abstract

The digital substrate – data, algorithms, infrastructure, platforms, applications – is being governed without adequate conceptual foundations. The *ability* and *legitimacy* required to govern this substrate, and to govern with it, are simultaneously misaligned, contested, and structurally absent. We introduce *digital statecraft* as the organising concept for this emerging field, arguing that 'digital' reconstitutes the statecraft question rather than merely extending its domain. The concept operates on two dimensions – statecraft *over* digital systems, concerning the authority and capacity of the state in relation to the digital substrate itself, and statecraft *with* digital systems, concerning the deployment of algorithmic tools as instruments of governing authority. And it rests on two foundational requirements, technical coherence and legitimate authority, that are genuinely in tension. We derive *ten principles of digital statecraft* from these foundations, each naming a condition whose absence produces an identifiable and structural governance failure: *public interest first*, *human-machine complementarity*, *governability by design*, *systemic coherence*, *hybrid institutions*, *adaptive governance*, *human centricity and civic agency*, *accountable and traceable authority, judgment across time*, and the *non-delegable core.* This article takes the state as the starting point, the institutional form that developed historically in response to the problem of effective and legitimate public governance, and the only current candidate for which the full set of legitimacy conditions – democratic mandate, constitutional accountability, legal continuity – is institutionally available. But the *digital statecraft* programme holds open a deeper question than just whether states can reform themselves: governing well in the algorithmic age may require rethinking the boundaries, scale, and affiliative basis of statehood itself.

---

[a] The Digital Statecraft Academy; [b] UCL Computer Science; [c] Department of Engineering, University of Cambridge; [d] University of Bristol; [e] Department for Digital, Culture, Media and Sport (DCMS), UK Government; [f] Department of Computer Science and Technology, University of Cambridge; [g] The Alan Turing Institute; [h] Swiss Federal Office of Energy; [i] Department of Mathematics, Imperial College London; [j] UK Statistics Authority; [k] Department of Labor, State of Maryland; [l] Digital Ethics Center, Yale University; [m] Innovorsa Group; [n] Dataswyft; [o] University of Tartu; [p] Prime Minister's Office, UAE; [q] Department of Legal Innovation, Hitotsubashi University; [r] EARTH Programme, University of Oxford; [s] James Madison University; [t] NYU GovLab; [u] The Data Tank.

*Correspondence: z.engin@digitalstatecraft.academy

Policy Significance Statement

Most digital governance asks whether systems are safe, fair and transparent. This paper asks a prior question: does the state have the ability and the legitimacy to govern the digital substrate and to govern with it? It shows why both are needed together. Ability without legitimacy is authoritarian efficiency; legitimacy without ability is institutional impotence. Ten principles turn this into a practical diagnostic, helping governments and policy makers locate where their arrangements fall short: in system design, oversight institutions, accountability chains across devolved and international boundaries, and the decisions that must keep a human answerable however capable algorithmic systems become. Together they offer a shared foundation for governing well in the algorithmic age.

# 1. Governing without settled authority

Algorithmic systems now govern public life in ways that were once the exclusive domain of human institutions: curating the information and public discourse citizens can access, determining eligibility for services, and mediating the relationship between citizens and the state. Whether they do so legitimately is a different question. What is missing from the governance arrangements that are proliferating around them is not more regulations, platform policies or international standards, but a coherent account of who has the authority and ability to govern these systems and to govern with them, on what basis, and with what legitimacy (Skorpen & Reinertsen, 2025; Yeung & Lodge, 2019; Lnenicka et al., 2026). This is not a technical problem dressed in political clothing, nor a political problem that better engineering can solve. It is both, and neither can be addressed without the other.

We propose *digital statecraft* as the organising concept for effective and legitimate public governance in the algorithmic age (Engin et al., 2025a). The term carries weight. First recorded in English around 1640 (OED, n.d.), roughly contemporaneous with the emergence of the modern state (Spruyt, 1994), *statecraft* has always presupposed the 'state' as the recognised actor with the authority, ability, and judgment to govern on behalf of a polity; and has always pointed beyond mere governance to something more demanding – the practical art of governing well (Caldwell, 1996; Bertelli and Peter, 2013). That art found its institutional expression in the industrial state: governing authority exercised through standardised rules, hierarchical processing, and bureaucratic machinery designed to produce predictable outputs from defined inputs (Weber, 2019). *Digital statecraft* does not inherit that model. The machinery of government is being reconstituted: algorithmic systems, non-state actors, and distributed authority are displacing the institutional arrangements on which industrial-era statecraft rested (Bratton, 2026; Fourcade & Gordon, 2020; Dunleavy & Margetts, 2025). That reconstitution raises the foundational question this paper holds open: whether the state remains the right primary actor for effective and legitimate public governance in the algorithmic age.

The choice of 'statecraft' rather than 'governance' is what makes that question answerable rather than merely askable. Digital governance describes the landscape of arrangements through which digital systems are governed (Charalabidis et al., 2022) – policies, regulations, international standards, algorithmic processes and the like. It is descriptive and pluralistic: existing, emerging, and proposed arrangements alike, however partial or inadequate. *Digital* *statecraft* makes a stronger and more specific claim: good public governance in the algorithmic age requires rebuilding the authority, judgment, and accountability that the digital transformation has put under pressure. *Digital statecraft* is the practice of doing so, governing well in conditions where the foundational presupposition of statecraft is itself under challenge, and where the very criteria against which 'governing well' is judged have been progressively narrowed to private sector metrics of efficiency and optimisation, displacing legitimacy and the public interest as the orienting purpose of governance (Mazzucato, 2021; Moore, 2013)[1]. The direction of travel is currently being set largely by private actors optimising for commercial metrics, not because public interest has been rejected as a value, but because no other actor has the mandate, capacity and occasion to assert it. Prioritising the public interest is therefore not one obligation among the state's several responsibilities in the algorithmic age; it is the responsibility that only the state, among current governing actors, is positioned to discharge.

**Principle 1 — Public Interest First**

*Digital systems deployed in governing contexts must serve the public interest as the orienting purpose of statecraft. This must be the primary standard against which governing arrangements are judged, and the corrective when that purpose is displaced by private capture or technical logic.*

At its core, *digital statecraft* requires two conditions simultaneously: the *ability* to act, and the *legitimacy* of governing arrangements through which it acts[2] (Table 1). *Ability* means the practical power to use, understand, inspect, contest, enforce, and adapt. *Legitimacy* means the basis on which authority can be justified and held to account. Political science has long distinguished formal authority from the practical capacity to act on it (Mann, 1984); political philosophy has long insisted that governing authority requires more than power to be sustained (Weber, 2019; Beetham, 2013; Habermas, 2015). *Digital statecraft* needs both integrated into a single analytical frame – separately specified, they leave the crisis unaddressed[3]. *Ability* without *legitimacy* is authoritarian efficiency, and *legitimacy* without *ability* is institutional impotence. What is needed is *statecraft*: the demanding practice of holding *ability* and *legitimacy* together under conditions where both are structurally contested.

[1] "Public interest" here draws on the existing public purpose/value tradition rather than advancing a novel account. Its scale (national, transnational, universal) and the process by which it is determined are contested within that tradition and are not resolved here; Section 3.1's treatment of affiliative legitimacy and jurisdictional dissolution bears directly on this question and is developed there.

[2] "Ability" is chosen over "capacity" – what institutions have – or "capability" – what systems can do; statecraft requires both and neither alone is sufficient. "Legitimacy" is chosen over "authority," which legitimacy confers rather than constitutes, and over "legality," which is a necessary but not sufficient condition for it. The choice of *ability* and *legitimacy* as the foundational pair equally requires justification. Other candidates from the statecraft tradition – resilience, sovereignty, representation, trust – are either derivative or constituent: resilience and sovereignty are conditions the two requirements jointly produce; representation and trust are legitimacy conditions of a specific kind rather than independent foundations. What distinguishes *ability* and *legitimacy* is structural: their absence produces identifiable failure modes in either direction, authoritarian efficiency and institutional impotence respectively, and no other pairing generates this bidirectional failure structure.

[3] For example, risk-based approaches (EU AI Act) calibrate obligations to potential harm without a theory of whether regulators can act on the obligations they impose; human-in-the-loop requirements assume that formal human presence satisfies accountability, a claim Section 2.2 disputes; technical standards (ISO/IEC 42001; NIST AI RMF) specify what compliant systems and processes should look like without addressing whether institutions can verify compliance in practice. Each paradigm addresses *ability* or *legitimacy* in some form; none integrates both into a single analytical frame.

***Table 1: Digital Statecraft Requirements and Dimensions***

| | Statecraft *over* digital systems | Statecraft *with* digital systems |
|---|---|---|
| **Ability** | Ability to act on the digital substrate itself | Ability to act through digital and algorithmic systems |
| **Legitimacy** | Basis on which authority over the substrate is justified | Basis on which authority exercised through digital and algorithmic systems is justified |

*Digital statecraft* has two dimensions that run through this article. Statecraft *over* digital systems concerns the ability and legitimacy of the state in relation to the digital substrate itself – the platforms, data, infrastructure, and algorithmic systems that constitute it. Statecraft *with* digital systems concerns the deliberate deployment of digital and algorithmic tools as instruments of governing authority. The paper uses '*over*' and '*with'* as shorthand for these two dimensions throughout.

The *over* dimension is the first site of destabilisation, and it operates on at least two levels. The first is a contest of power: large technology companies now exercise functions that were historically state or civic society functions (Van Dijck et al., 2018; Rahman, 2018; Zuboff, 2023; Acemoglu & Johnson, 2023; Kalema, 2024; Gohwong, 2026) – setting norms of public communication, controlling digital infrastructure, making consequential decisions about access to employment, credit, and public services – largely without democratic mandate and with accountability structures that bear no resemblance to those we apply to public institutions. This contest of power is also a realpolitik one: states increasingly have to choose which hegemon's data and compute infrastructure to depend on, a strategic alignment question that sits alongside, and often overrides, the accountability question. The second is a deficit of technical capacity: even where states hold formal authority, they may lack the technical means to exercise it meaningfully; and this deficit is not uniformly distributed (Nikiforova et al., 2026). States in the Global South frequently face a compounded version: formal authority over digital systems that depend on infrastructure they do not control, regulatory capacity that is structurally outmatched by the technical and legal resources of the actors they are attempting to govern, and populations subject to algorithmic systems whose governance arrangements were designed elsewhere and remain accountable elsewhere. That compound deficit is sharpest where it intersects with a second structural problem: algorithmic systems are often opaque by design and resistant to the kind of inspection that regulatory oversight assumes (Pasquale, 2015; Wachter et al., 2017). A regulator can mandate transparency and still be unable to assess whether a system is genuinely transparent; and even where systems are transparent, non-deterministic outputs can exceed any inspection regime anticipates, and interpretability requires both technical tools and organisational capability that governing institutions rarely possess simultaneously. Governing authority over digital systems can therefore be formally asserted and practically unverifiable. The state can no longer presuppose either its authority over this terrain or the legitimacy of its claim to govern it.

The *with* dimension is the second site of destabilisation. Algorithmic processes offer genuinely expanded governing capacity – more responsive public services, better evidenced policy decisions, greater reach and efficiency at scale (Engin & Treleaven, 2019; Nikiforova et al., 2026). These are not marginal gains: they represent a qualitative shift in what governing institutions can do on behalf of the people they serve. That shift runs in two directions:

governing institutions can do more, but so can the individuals they serve. Digital systems can make governing processes legible, expand access to public services, and equip individuals to navigate and engage with the systems shaping their lives. But two problems arise together. Public institutions often lack the technical knowledge, infrastructure, and resources to deploy these capabilities effectively, while private sector actors do so readily – creating an asymmetry that shifts governing functions away from accountable public institutions. And when institutions exercise governing authority through algorithmic processes, this raises accountability and responsibility requirements that efficiency gains alone cannot resolve.

Of all the ways "digital" destabilises the statecraft presupposition, *agentic AI* tests the *craft* claim most severely. Agentic systems do not merely execute pre-specified rules – they plan, adapt to context, and pursue goals across sequences of actions (Acharya et al., 2025; Suleyman, 2023; Hughes et al., 2025) in ways that begin to resemble *craft* rather than procedure. A *craft* requires learned skill, judgment, and the kind of practical wisdom – *phronesis* rather than *techne* – that cannot be fully reduced to rules or optimised away (Aristotle, Nicomachean Ethics, Book VI). This is not only a philosophical claim: the craft of translating evidence into policy decisions depends on judgments of timing, audience, feasibility, and trust that cannot be derived from the evidence itself or specified as a procedure in advance. The same evidence, presented at the wrong moment, to the wrong audience, or without an accompanying account of feasibility, fails to inform the decision it was meant to serve regardless of its technical quality. This practical dimension of statecraft, arguably the clearest surviving instance of phronesis in contemporary governing institutions, is what agentic delegation puts under the greatest strain. If statecraft is irreducibly a craft, the question becomes whether that craft can be delegated to systems that exhibit something structurally similar to practical judgment. This article's answer is not that agentic systems are incapable – that debate is unresolved across technical, philosophical, and political registers. It is that statecraft involves accountability as a constitutive element (Bovens, 2007). The Non-Delegable Core (NDC), developed in Section 3.3., is a claim about the accountability structure that must surround any decision that exercises governing authority, not about the quality of algorithmic output. Agentic AI makes that claim urgent rather than merely theoretical.

The tension between what algorithmic systems make possible and what legitimate governance requires cannot be resolved by choosing one side. It is the central condition of *digital statecraft*, which must be actively governed through the deliberate design of arrangements that harness machine capability without displacing human authority (Pasquale, 2020; Floridi et al., 2018).

> **Principle 2 — Human-Machine Complementarity**
>
> *Governing well in the algorithmic age requires the deliberate design and active stewardship of hybrid environments where human judgment and machine capability are co-required – neither automating without governing nor governing without capability.*

Where this condition goes unmet, institutions default to one side or the other by omission – full automation with no route for human judgment to intervene, or unaided human process with no algorithmic support to catch what humans cannot see at scale. The public bears the cost of a complementarity that was never attempted. That cost is not only a governance failure but a public interest one (Principle 1): where a capable system would materially improve outcomes for those the state serves, declining to deploy it also fails the public it exists to serve (Watson et al., 2024).

The choice of ‘digital’ rather than ‘AI’, ‘algorithmic’, or ‘agentic’ statecraft is deliberate. AI operates within and depends on a broader substrate – data, connectivity, cloud, compute, hardware supply chains, etc. – that is politically consequential, unequally distributed, and often privately controlled (van Kranenburg, 2024; Pohle & Thiel, 2020). *Digital statecraft* insists on the full stack as the legitimate scope of the statecraft inquiry (Bratton, 2026; Engin et al., 2025b). A state that reasserts authority over platforms but deploys opaque algorithmic systems against its own citizens has not resolved the problem; it has centralised it. A state that insists on human judgment in its own processes but cedes authority over digital infrastructure to private actors has not resolved it either; it has externalised it. And a state that governs with algorithmic systems while remaining dependent on privately controlled compute, data, and connectivity has secured neither *legitimacy* nor *ability.* All requirements and dimensions must be addressed together: that is what makes *digital statecraft* a distinct and demanding intellectual project, and what distinguishes it from the governance frameworks that have so far proven inadequate to the challenge.

The sections that follow develop these foundational requirements. The sequencing is analytical, not hierarchical: *ability* (Section 2) and *legitimacy* (Section 3) are co-required. Ten principles are derived across the article, each naming a condition whose absence produces a structural governance failure. The article's most distinctive claim is that the tension between the two requirements has a limit, the Non-Delegable Core, where legitimacy conditions converge into a structural requirement that technical capability cannot satisfy. Section 4 holds open the deeper question this raises: not only whether states can rebuild the *ability* and *legitimacy* conditions the algorithmic age requires, but whether the boundaries and basis of statehood itself are being reconstituted in the process – a question this paper establishes the foundations for asking rigorously, rather than answers.

## 2. What *ability* requires: technical coherence

The *ability* to govern well in the algorithmic age is not simply a matter of more technical capacity, better regulatory tools, or more technically literate public servants – though all of these matter. It requires that the systems through which statecraft is exercised, and the institutions exercising it, are functionally coherent across the full architecture of public governance. *Ability* is not an institutional property in the abstract; it is the precondition for the quality of public life. Where it is absent, decisions affecting access to employment, housing, healthcare, and justice are made by systems that accountable institutions cannot verify, contest, or correct. We call this structural requirement *technical coherence*. It is not purely technical: the *ability* conditions developed here are conditions for governing well, not merely governing effectively, and legitimacy constraints already shape what *ability* requires and what counts as having it.

*Technical coherence* operates at two levels. At the systems level, individual systems must be designed for governability, and the full-stack architecture must be treated as a coherent whole (2.1). At the institutional level, new hybrid oversight structures are needed, responsive to algorithmic tempo and behaviour (2.2). Neither is sufficient alone, and both share a vulnerability: each is typically specified against a particular configuration of human-machine relationships, and both must hold as that configuration shifts (2.3).

## 2.1. The systems level: individual systems and full-stack architecture

The gap between technical compliance and governing capacity is where the *ability* to govern well is most routinely lost. A system can be technically transparent but practically incomprehensible to the regulator overseeing it; it can disclose decision logic without enabling independent verification. Compliance is satisfied; governing capacity is not. The pattern is not new: Basel II capital adequacy frameworks required internationally active banks to disclose internal risk models and demonstrate compliance, yet regulators systematically lacked the technical capacity to assess whether those models were doing what they claimed – a governing gap whose consequences became visible only in the 2008 financial crisis (Daníelsson, 2002; Power, 2007). Closing that gap requires reframing the relevant technical properties such as safety, fairness, transparency, accountability, and explainability – not only as properties that well-designed systems should exhibit, but also as conditions specifying what governing institutions must be able to do (Koshiyama & Engin, 2019; Dwivedi et al., 2021). That reframing shifts the question from what systems exhibit to what institutions can actually do with them. Table 2 summarises six *ability* conditions, each naming what governing institutions must be able to do and what is at stake for public life when it is unmet.

***Table 2: Systems-level ability conditions***

| ***Ability* condition** | **Governing gap and what is at stake** |
|---|---|
| **Use** – deploy digital systems as governing instruments | Without deployment capacity, public governing authority migrates to private actors by default, not by democratic choice. Public services become less responsive and accessible, with costs falling disproportionately on those least able to seek alternatives. Public interest obligations and contestability rights disappear. |
| **Understand** – comprehend what systems are doing and why | Post-hoc explanation satisfies disclosure without enabling comprehension (Arrieta et al., 2020; Lipton, 2018; Wachter et al., 2017; Janssen et al., 2022). People are governed by systems no accountable authority genuinely understands. |
| **Inspect** – independently verify system behaviour | Audit methodologies exist but are not deployed by governing institutions; agentic systems resist pre-specified audit frameworks (Raji et al., 2020; Barocas, Hardt & Narayanan, 2023; Mökander et al., 2024). Bias and failure accumulate without correction. |
| **Contest** – challenge operator claims | Formal contestation rights exist without the ability to exercise them (Diakopoulos, 2016; Mulligan & Bamberger, 2019). Governing decisions are practically unchallengeable regardless of formal rights. |
| **Enforce** – verify compliance, detect evasion | Whether standards produce the behaviours they mandate is a structurally harder question than current frameworks answer (ISO/IEC 42001; NIST AI RMF). Formal authority exists over systems whose actual behaviour cannot be verified (Pasquale, 2020). |
| **Adapt** – anticipate and update governing capacity as systems evolve | Frameworks calibrated to yesterday's systems govern today's in name only. Governing institutions that cannot anticipate configuration shifts are structurally reactive, always catching up to failures already experienced by those subject to them (Engin, 2026; METI, 2021). |

In practice, deployment capacity is tested earliest at the point of commissioning. Institutions that cannot specify governability requirements in procurement, cannot evaluate supplier claims against them, or challenge suppliers when systems fail to meet them have already lost the *ability* condition before a system is ever deployed. This is a capability gap distinct from technical literacy: it is the institutional capacity to ask better questions of the market, and it shapes every one of the six conditions in Table 2 upstream of deployment. All six conditions follow the same logic downstream: none is a property that can be added after the fact. Each must be built into systems from the outset and must hold not only at the moment of deployment but across the full life of a system and the full range of configurations in which it will operate – extending the logic that 'privacy by design' and 'security by design' established in adjacent domains (Cavoukian, 2009). Opacity, resistance to audit, and contestation barriers are not neutral design choices. They are conditions that structurally undermine the *ability* to govern and impose direct costs on the people subject to ungoverned systems.

**Principle 3 — Governability by Design**

*Systems deployed in public governance contexts must be designed for governability, contestability, and functional interpretability from the outset. These must be core design requirements operable across the full life of the system and the full range of configurations in which it might operate. They should not simply be constraints imposed after deployment.*

Governing individual systems well is necessary but not sufficient. The digital substrate of states is not a collection of independent systems but an architecture of mutually enabling layers: foundational infrastructure, data capabilities, algorithmic processes, and citizen-facing services, whose enabling relationships run in both directions across the stack (Meadows, 2008; Engin et al., 2025b). Governing failures at any point propagate across the whole. An analogy with modular software architecture is instructive: well-designed modular systems achieve coherence not through unified control but through clear interfaces between independent components that let information and accountability flow across boundaries without gaps (Parnas, 1972). Governing well therefore requires the full stack to be treated as a coherent governance object in precisely this sense – not a monolith under unified control, but an architecture in which each component operates independently while exchanging information and accountability across boundaries without governance gaps opening at the seams. End-to-end accountability must be traceable from the hardware layer through to the decisions it enables.

The *over*/*with* distinction clarifies what this demands in practice. On the *over*-axis, the ability conditions concern what governing institutions must be able to do in relation to private actors exercising quasi-public functions; full-stack coherence means asserting meaningful sovereignty over the layers governing systems depend on, not merely regulating their surface. On the *with*-axis, the ability conditions concern the accountability structures needed when the state itself deploys algorithmic systems; full-stack coherence means ensuring the integrity of the state's own internal architecture so that governing through algorithmic systems is genuinely accountable end-to-end. What differs is who bears the requirement and against whom it must be exercisable. On the *with*-axis, this carries a further implication: the ability conditions are not only requirements on what institutions can do but legitimacy constraints on what governing through algorithmic systems can permissibly look like, a distinction Section 3 develops in full.

Systemic coherence is a claim about accountability, not control. It does not mean unified governance. Legitimacy conditions (Section 3) can generate requirements for designed separation: purpose limitation constrains integration, authorisation chains derived from different legal bases cannot be merged without democratic mandate, and highly integrated architectures can undermine the individual contestability. Designed separation is therefore not a failure of systemic coherence, but an expression of it. What coherence requires is the elimination of ungoverned fragmentation: gaps where accountability dissolves by default rather than by design. A governing architecture can be fully coherent while deliberately distributed; what it cannot be is accidentally ungoverned.

**Principle 4 — Systemic Coherence**

*Digital infrastructure, data, algorithmic systems, and platforms must be governed as a coherent whole – maintaining end-to-end accountability traceable from the hardware layer through to the decisions it enables. Systemic coherence avoids the perils of both centralisation, with its top down accountability, and ungoverned fragmentation, which undermines accountability.*

## 2.2. The institutional dimension: from periodic oversight to hybrid institutions

Section 2.1 concerned properties of what is being governed. This subsection concerns the governing structures themselves – the institutional forms through which technical coherence is exercised in practice. Institutional design failures are not reducible to systems failures: an institution can govern a well-designed, fully interpretable system and still fail to exercise meaningful oversight if its own architecture is mismatched to what algorithmic governance demands. Those demands go beyond tempo and scale, though the speed and volume of algorithmic decision-making already exceed what industrial-era oversight was designed to handle (Citron, 2007). They also concern opacity and emergent risk, the infrastructural character of algorithmic systems, adaptive capability, and distributed agency amongst others – each requiring different institutional forms, not faster versions of existing ones. Table 3 maps these mismatches and their design consequences; when they persist, the costs fall on the people whose lives are shaped by decisions institutions cannot adequately oversee.

***Table 3: Institutional-level ability considerations***

| Dimension | What industrial-era oversight assumed | What algorithmic age oversight requires |
|---|---|---|
| **Tempo and scale** (Citron, 2007; Zarsky, 2016) | Discrete decisions at human pace, amenable to periodic review. | Hybrid institutions operating continuously across overlapping, simultaneous decisions at machine scale, with algorithmic systems surfacing those requiring human judgment. |
| **Opacity and risk** (Pasquale, 2015; Leveson, 2016) | Inspectable objects with discrete, traceable failure modes. | Interpretability tools and automated audit combined with human ability to assess outputs – neither is sufficient alone. |
| **Algorithmic systems as infrastructure** (Plantin et al., 2018; Rahman, 2018) | A regulated entity that oversight stands outside and inspects. | Accountability architectures built into the substrate itself; human authority constitutive at design and governance level. |
| **Adaptive capability** (Calo, 2017; Acharya et al., 2025; Engin, 2026; METI, 2021) | A stable system whose decision logic can be assessed at a point in time. | Algorithmic systems tracking configuration shifts; human institutions retaining judgment about when recalibration is required. |
| **Distributed agency** (Mittelstadt et al., 2016; Floridi et al., 2018) | A singular accountable actor who can be held responsible. | Machine capability tracing agency across institutional boundaries; human institutions accountable for what happens at them. |

Institutional innovation is emerging – but unevenly, partially, and without the architectural coherence required (Haug et al., 2024). Upstream assessment requirements, transparency registers, and expanded audit mandates address individual mismatches in isolation rather than as components of an integrated oversight architecture (Mulgan, 2017). A 2025 national study of generative AI use in the UK's Research Excellence Framework illustrates the pattern: institutions adopted the technology disparately and largely ungoverned, with no body positioned to oversee the pattern that adoption produced across the sector (Benites et al., 2026), prompting calls for a dedicated national AI governance framework and shared infrastructure to prevent inequitable practice (Watermeyer et al., 2025). No jurisdiction has yet developed oversight arrangements that fully address the governance-from-within problem: governing systems that have become the substrate of governing itself.

The deeper problem is architectural and constitutional. The five mismatches in Table 3 converge on a single challenge: industrial-era governance rested on periodic legitimacy – authority granted, exercised, and renewed at fixed intervals. Algorithmic systems break that presupposition: they transform the nature and consequences of governing decisions continuously, in ways that periodic accountability cannot track or correct (Hildebrandt, 2015; Cohen, 2019). A parliament's role is to hold the executive accountable for ensuring effective oversight of algorithmic systems exists – but if parliament only does this once a session, it is reviewing a historical record, not exercising scalable oversight. Structures to sustain *legitimacy* must be functional in real time – through continuous scrutiny, contestability, and capacity for correction (WEF, 2018; OECD, 2021).

Both the operational mismatches and the constitutional challenge require new institutional forms, hybrid in character: combining machine capability with human judgment rather than substituting one for the other. Oversight forms cannot operate at purely human scale but cannot dispense with the judgment that bears accountability – oversight without it is simulation. Accountability forms require automated monitoring to surface what matters, but the accountability judgments themselves must rest with human institutions. This is not what 'human-in-the-loop' arrangements typically deliver: nominal oversight, where a human is formally positioned to review algorithmic outputs without the capacity or authority to genuinely contest them, satisfies the form of accountability without its substance. Citizen-facing forms can deploy digital tools to expand access and agency, but that agency remains irreducibly human (Noveck, 2015). In each case the design challenge is to build hybrid environments where machine capability is required for scale, monitoring, and adaptation, and human judgment is required for accountability, legitimacy, and contestation – neither substitutable for the other.

**Principle 5 — Hybrid Institutions**

*Governing at algorithmic speed and scale requires new institutional forms that sustain legitimacy continuously. These forms are hybrid: machine capability extends the reach and resolution of human judgment; human authority remains constitutive where accountability requires it. They cannot be derived from industrial-era institutions alone, nor achieved by human institutions or algorithmic systems operating alone*[4]*.*

## 2.3. The dynamic dimension of technical coherence

The requirements developed in Sections 2.1. and 2.2. share a vulnerability: both are typically specified against a particular configuration of human-machine relationships, and both can become misaligned with the governing reality as that configuration shifts, without any failure in the systems or institutions themselves. A system designed for governability in one configuration may shed that property as it scales, retrains, or is redeployed. A hybrid institution designed to govern one distribution of authority and accountability may be entirely misaligned with what that distribution has become. In the algorithmic age this is compounded: systems and institutional orders are produced together, each shaping the conditions under which the other operates (Bratton, 2026; Luna-Reyes & Gil-Garcia, 2014), so configuration shifts in one propagate into the other in ways neither can track alone. This is already documented within algorithmic systems themselves, where changes in data collection practices can silently degrade model performance through data or concept drift (Widmer & Kubat, 1996; Lu et al., 2018) – governing failures neither data governance arrangements nor algorithmic oversight arrangements were designed to detect.

Technical coherence requires that both hold across shifting configurations. Configuration means more than temporal change: the same system deployed at a different scale, institutional setting, or oversight structure constitutes a different configuration even where no technological change has occurred (Selbst et al., 2019; Engin & Hand, 2026). What counts as governable, auditable, or contestable in one configuration may not hold in another. The *over/with* boundary itself shifts: as states acquire capability, functions that were once on the *over*-axis migrate to the *with*-axis. A fraud-detection system procured from a private supplier sits on the *over*-axis; once the state builds equivalent capability and operates it directly, the function moves to the *with*-axis, and accountability shifts from oversight of a supplier to the state's own internal accountability for a decision it now makes itself. Governance arrangements that do not track this migration will apply *over*-axis frameworks to *with*-axis problems, or vice versa.

That mismatch is invisible to static frameworks and consequential for both *ability* and *legitimacy* alike – precisely where the two foundational requirements are most likely to demand active governing judgment rather than the application of settled frameworks. Static frameworks fail here (Brownsword, 2008): the failure is architectural, not a matter of being out of date. Adaptability is therefore not a feature of well-designed systems or well-designed institutions considered separately. It is an essential property of technical coherence itself. In the algorithmic age, deployment is the starting point of governing responsibility, not its endpoint:

[4] Design alone does not guarantee this in practice: even well-designed hybrid institutions depend on a governing culture willing to exercise contestation and judgment rather than deferring to algorithmic outputs by default – a condition this article identifies as necessary but does not itself specify.

the directions a system takes afterward – through use, retraining, and contextual variation – may be entirely unforeseen, and governing arrangements that cannot track and respond to those shifts are misaligned with the systems they claim to govern.

> **Principle 6 — Adaptive Governance**
>
> *Governance arrangements must track the actual distribution of authority and accountability as human-machine configurations shift, and recalibrate accordingly. Static frameworks fail structurally, and adaptability is a defining feature of technical coherence.*

Section 2 has developed what *ability* requires: systems designed for governability and systemic coherence across the full stack; institutions capable of operating at algorithmic tempo in hybrid form; and governance arrangements that hold both together across shifting human-machine configurations. These are necessary conditions, not sufficient ones. A governance architecture that satisfies every requirement above can still be illegitimate: if those exercising authority cannot justify it to those subject to it, cannot be held accountable for it, and cannot be corrected when they exercise it wrongly, technical coherence has produced not digital statecraft but authoritarian efficiency.

## 3. What *legitimacy* requires: justifiable authority in the algorithmic age

Serving the public interest is necessary but not sufficient. *Legitimacy*, as established in Section 1, is the basis on which authority can be justified and held to account. Justification and accountability both presuppose a party to whom authority is answerable – capable of assessing whether it has been exercised well, of being wronged by its failure, and of holding it to account (Bovens, 2007; Beetham, 2013). That party cannot be an aggregate metric or a technical property of a system; it can only be a rights-bearing subject. Legitimate governing authority therefore exists for a purpose: to serve human beings as rights-bearing persons, citizens, and the ultimate subjects of the power exercised over them – and because they are its addressee, they are also its test: a governing arrangement is assessed by whether it succeeds for the people subject to it.

Principle 7 is deliberately framed in terms of the 'human' rather than the 'citizen'. A great deal of algorithmic governing authority is now exercised not by states over 'citizens' but by platforms over 'users' – a relationship citizenship was never built to describe, offering no membership, mandate, or representation in return, only a contract. Human-centricity names the wider category *legitimacy* must answer to; civic agency names the specific capacities a polity owes those within its own authority.

> **Principle 7 — Human Centricity and Civic Agency**
>
> *The human being is both the measure and the purpose of digital statecraft. Every governing arrangement must sustain the capacity of those subject to it to understand, contest, and shape the systems that govern them – substantively, not formally; collectively, not only individually. The ambition is a polity empowered to govern itself through and alongside digital systems.*

The conditions that follow elaborate this single commitment. Unlike Section 2's *ability* conditions, which specify what institutions must be able to do, these specify limits on what governing with digital systems can permissibly look like – limits technical capability cannot override. Three follow: the institutional structure of legitimate authority, requiring accountable human agents traceable through unbroken chains of authorisation (3.1); judgment across time, requiring authority held with practical wisdom and stewardship for consequences beyond the present (3.2); and the Non-Delegable Core, the limit where these conditions converge into a requirement no degree of technical capability can satisfy (3.3).

## 3.1. From accountability to authorisation: the institutional conditions

Four requirements specify what the institutional structure of legitimate authority demands. They are analytically distinct but structurally interdependent: each presupposes the others, and all four fail together wherever the chain of legitimate authority is broken.

*Accountability* requires more than answerability. An algorithmic system can be made answerable – its outputs logged, its decision logic inspected, its performance audited. Accountability requires a human agent or institution that bears responsibility for the exercise of authority and can be sanctioned, removed, or overruled (Pettit, 1997; Bovens, 2007; Hildebrandt, 2020). That agent must be human not because human judgment is superior – that debate is genuinely unresolved – but because authorisation is structurally a human relationship (Rouvroy & Berns, 2013; Hart, 1961). *Traceable authority* is accountability's precondition: bearing responsibility presupposes that the authority being exercised is genuinely one's own to bear, traceable through an unbroken chain of authorisation connecting governing decisions to those they represent, running through institutions and infrastructure under legitimate control. Where that chain is broken, responsibility cannot be borne – only performed. *Consent* is the originating claim of legitimate authority: it derives from those subject to it, not merely operates over them accountably, and requires genuine civic agency – the practical capacity to understand, engage with, and shape the arrangements through which governing authority is exercised (Coeckelbergh, 2023). Consent without that capacity is the appearance of consent produced by the absence of alternatives. Civic agency is also collective: a polity's capacity to understand, contest, and shape governing systems is a distinct legitimacy condition, not derivable from aggregating individual contestation rights. *Revisability* is the mechanism holding the other three together: governing decisions must be contestable, and governance arrangements must include mechanisms for identifying and correcting errors (Bovens, 2007), reaching a human authority capable of bearing responsibility. Recursive algorithmic review – AI System A reviewing AI System B – extends the authorisation chain without terminating it in a human decision-maker, adding complexity without adding human comprehension or responsibility.

In industrial-era governance, these four requirements had settled institutional form. *Accountability* terminated in a named human official – a minister bearing personal responsibility for every decision made under their authority, answerable to parliament and removable for its exercise. *Traceable authority* ran through a legible chain – civil servant, minister, parliament, electorate – each link named and in principle contestable. *Consent* was mediated through representative institutions that, however imperfectly, grounded governing authority in democratic mandate. *Revisability* operated through legal challenge, parliamentary scrutiny, and administrative appeal – slow and incomplete, but present. These mechanisms rarely operated without friction, but their institutional form was clear, their accountability assignments were nameable, and their failure modes were visible. Underpinning all four was a rarely explicit territorial assumption: that those subject to governing authority were affiliated with the institutions exercising it, through citizenship, residency, shared history, and belonging to a bounded polity. Legitimacy was not only procedural but affiliative (Miller, 1995; Anderson, 1983; Renan, 1882) – the state's authority was recognised partly because people derived identity, protection, and meaning from membership in it[5]. That assumption is itself now under pressure: Estonia's e-Residency programme asks whether participation in a state must be tied to geography, and the European Union exercises statecraft functions across borders without being a state in the traditional sense. These are not merely jurisdictional complications – they signal that the affiliative basis of legitimate authority cannot be assumed when digital systems are reconstituting the boundaries of belonging itself.

Digital transformation disrupts all four requirements, and their territorial assumptions, differently on each axis. On the *over*-axis, the failure is displacement. Platforms exercise governing functions without the institutional forms through which the four requirements are satisfied: no human agent bears responsibility for governing consequences as a public official does, authority is not traceable to democratic mandate, those subject to platform governance ('users' rather than 'citizens') did not authorise it and cannot meaningfully shape it, and revisability reaches no human authority bearing public responsibility. Because that population is defined by use rather than membership, civic agency (Principle 7) cannot remain gated by citizenship either – human rights already extend to personhood rather than membership, and civic agency must extend at least as far in practice.

On the *with*-axis, the failure is internal dissolution. When agentic systems sequence consequential decisions about welfare, immigration, or child protection, accountability dissolves across designers, procurers, deployers, and overseers – none bearing responsibility in the ministerial sense, all pointing elsewhere (Grimmelikhuijsen & Meijer, 2022). Traceable authority breaks at the infrastructure layer: a state deploying AI-based determinations on infrastructure it does not control – whether foreign-owned, domestically-owned, or privately operated within national borders – and trained on data whose provenance it cannot verify, has not secured its chain of authorisation to its constitutional basis (Pohle & Thiel, 2020). Digital sovereignty is therefore a legitimacy condition – one of enforceable accountability over infrastructure, not ownership or territorial control (Bratton, 2026). And consent requires that human subjects can genuinely understand and contest the algorithmic arrangements governing them – a condition that opaque, privately procured systems resist regardless of where they are based. Where systems cross borders, all four failures compound – most acutely in Global South contexts, where infrastructure dependency and regulatory asymmetry further remove affected populations from the political systems governing the systems that govern them. Jurisdictional dissolution is a compound legitimacy failure that establishes international cooperation as a requirement, not a strategic preference.

[5] That reciprocity historically had a fiscal dimension as well as an identity-based one: the link between taxation and representation is a foundational claim in the fiscal-sociology tradition of state formation (e.g. Tilly, 1992), where the obligation to tax was met with an obligation of representation and accountability in return. Digital platforms complicate this further: value is frequently extracted from users in one jurisdiction while profit is booked and taxed in another, so even where a user's data or attention originates within a state's territory, no fiscal relationship analogous to citizenship's reciprocal claim need exist between the user and the platform governing them. This is a further destabilisation of affiliative legitimacy beyond the geographic and formal-membership dimensions discussed above, and is not developed further here.

The account above addresses jurisdictional dissolution across state borders: authority breaks when infrastructure, data, or governing functions cross into jurisdictions the affected population cannot hold accountable. A parallel failure mode operates inside a single state's own constitutional structure – one the industrial-era account of traceability did not need to address, because the chain it described (civil servant, minister, parliament, electorate) presupposed a single tier of democratic mandate. States with devolved, federal, or multi-level constitutional arrangements have several such chains, each legitimate within its own competence and capable of diverging from the others in the digital and data policy it authorises (Moon & Evans, 2017).

Algorithmic systems rarely respect this structure. A system designed, procured, and authorised at one level of government – usually the level with the greatest capacity to build it – is frequently deployed against a population whose relevant public services are governed by a different, constitutionally distinct authority. Where that is so, the traceable authority condition cannot be satisfied by tracing the chain to the level that built the system; it must trace to the level that holds the constitutional mandate over the domain. A national system deployed across a devolved health, education, or justice domain is not merely a coordination problem. In UK practice, this fault line is already visible: National Health Service (NHS) delivery is devolved to Scotland and Wales, but the data protection regime and medical device regulation governing AI systems used within those services remain reserved to the UK Parliament (Scottish Parliament Information Centre, 2024; Senedd Research, 2024). The authorisation chain for such a system therefore does not run through the level of government holding democratic mandate over the service it operates within: authority is exercised over a domain by a chain of authorisation that does not run through the body constitutionally mandated for it. This distinguishes two dimensions of traceability. *Horizontal traceability* concerns whether authority can be traced across state boundaries to a body the affected population can hold accountable. *Vertical traceability* concerns whether it can be traced through the correct tier of a single state's constitutional structure to the level of government holding democratic mandate over the domain it governs. Both are required: a system can pass the horizontal test – procured and controlled entirely within the state – while failing the vertical one, authorised by a level lacking constitutional competence over the domain it governs. Multi-level and devolved states should expect traceability to generate intra-state design requirements as well as cross-border ones: systems whose deployment crosses a devolution boundary need an authorisation architecture built to match, not a single national mandate assumed to cover it.

**Principle 8 — Accountable and Traceable Authority**

*Governing authority is legitimate only where it is accountable – borne by identifiable humans who can be sanctioned or removed – and traceable through an unbroken chain of authorisation to those it represents. The two fail together, whether through algorithmic delegation without human presence, dependency on infrastructure outside the state's control, platform capture of governing functions, or jurisdictional dissolution across or within state borders.*

## 3.2. Judgment across time: *legitimacy* as stewardship

The conditions established in 3.1 concern the institutional structure through which legitimate authority is exercised in the present: the accountability chain, the infrastructure it flows through, the consent of those subject to it. But governing well has always required something more than getting that structure right at a moment in time. It requires judgment – the practical wisdom that statecraft demands (Aristotle, Nicomachean Ethics, Book VI; Caldwell, 1996) – and judgment is irreducibly temporal: it applies principle to circumstances that could not have been fully anticipated, and it must be held accountable for consequences extending beyond the act of deciding. Judgment is a *legitimacy* condition, not merely an *ability* condition, because of this temporal structure. A system can simulate the outputs of judgment, but simulating an output is not the same as exercising judgment: the legitimacy condition concerns the accountability structure within which the output is produced, not the output's quality (Bovens, 2007; Pettit, 1997). And accountability, as Section 3.1 established, requires human agents capable of bearing responsibility across time, including for consequences they could not have fully foreseen.

Three arguments establish why the temporal dimension of governing authority generates distinct legitimacy requirements beyond those Section 3.1 already demands. The first is the *foreclosure argument*. Governing decisions made now can undermine the legitimacy conditions for future governance by locking in technical architectures, concentrating power, or closing off contestation pathways in ways that future publics cannot reverse (Collingridge, 1980). An algorithmic system embedded in critical public infrastructure – a benefits determination system, a policing risk-assessment platform, a public health triage tool – creates path dependencies that shape what future governing arrangements are practically possible. Where those dependencies are created without democratic mandate, or in ways resistant to future modification, they violate the traceable authority condition prospectively: future publics subject to those systems cannot trace the authority to themselves, because the decisions that created the dependency were made without their participation and cannot readily be reversed.

The second is the *accumulation argument*. Algorithmic systems governing public life accumulate effects over time in ways that individual contestation cannot address and that no single decision-maker intended (O'Neil, 2016; Eubanks, 2018). A welfare assessment system making individually contestable decisions may systematically disadvantage particular groups across repeated decisions, visible only in aggregate. A predictive policing system recalibrating continuously on its own outputs can entrench geographic and demographic patterns of enforcement that compound across years. Recommender systems shaping access to public information can progressively narrow the epistemic commons democratic deliberation depends on. Each individual decision may satisfy the accountability and traceability conditions of Section 3.1; the accumulated pattern satisfies none of them (Hand, 2026), producing effects – consequential, coercive, and population-wide – that fall entirely outside the institutional structures through which legitimate authority operates. Where accumulated effects systematically disadvantage particular groups across millions of decisions, the distributional consequences of governing through algorithmic systems become a legitimacy condition in their own right, not reducible to individual contestation rights or aggregate efficiency gains (Moore, 2013). This is not merely a fairness problem, though it is that: governing effects that cannot be traced, contested, or held accountable are not legitimate, regardless of the technical properties of the systems producing them. The accumulated effects must be visible, assessable, and correctable as a legitimacy condition, not merely a technical one. The institutional forms capable of doing so are still at the early stage of development and geographic deployment.

The third is the *generational consent argument*. Those who will bear the consequences of current digital governance decisions have not consented to them and cannot yet participate in making them (Rawls, 1971; Jonas, 1984). If governing authority must be traceable to those it represents, then the accountability and traceability conditions must apply prospectively as well as retrospectively. A governing architecture optimised entirely against present preferences – even one that satisfies every institutional condition Section 3.1 derives – fails a legitimacy condition no less real for being temporal: future persons subject to today's governing decisions are governed without their consent and without accountability to them. The disposition the judgment condition requires is stewardship: holding authority in trust for those represented, including those not yet present to claim it.

Together, these three arguments establish stewardship not as a peripheral commitment but as the governing disposition the *legitimacy* conditions themselves require. Accountability without stewardship expires at the boundary of the present; traceable authority without stewardship is a chain that terminates at the living. The judgment condition extends both across time.

> **Principle 9 — Judgment Across Time**
> *Legitimacy extends across time. A decision that forecloses future contestation is not fully legitimate, and accumulated effects must be assessed as a legitimacy question. Digital systems are public assets, held in trust for present and future publics alike.*

## 3.3. The Non-Delegable Core: where the tension inverts

For most governing decisions, the tension between *technical coherence* and *legitimate authority* is navigable through practical wisdom: systems can be designed for accountable decision-making, infrastructure dependencies mitigated through regulatory design, accumulated effects monitored by hybrid institutions. But navigability has a limit, defined not by technical inadequacy but by the structure of legitimate authority itself. There is a class of decisions – those involving fundamental rights[6], coercive power, irreversible consequences, or contextual moral judgment that cannot be fully specified in advance – where the *legitimacy* conditions converge on a single structural requirement that algorithmic systems cannot satisfy. Accountability requires a human agent constitutively present in the act of deciding, not supervising from a distance or reviewing outputs after the fact. Traceable authority requires the chain of authorisation to terminate in a human decision-maker: authorisation is structurally a human relationship (Hart, 1961), and a chain terminating in an algorithmic system has not been traced to anyone. Judgment across time requires agents capable of being held accountable for consequences that could not have been fully anticipated, including foreclosure, accumulation, and generational impacts. Together these convergent requirements define the Non-Delegable Core (NDC) (Engin & Hand, 2025): the class of governing decisions whose legitimacy depends on human authority being constitutively present in the act of deciding – not because algorithmic systems cannot produce good outputs, but because the accountability structure that makes such decisions legitimate cannot be borne by systems that cannot bear responsibility.

[6] The paper does not advance a novel account of which rights qualify as fundamental. The claim rests on existing constitutional and human rights traditions, those rights whose violation cannot be adequately remedied after the fact and whose determination therefore requires the full accountability structure that Section 3.1 derives. Contextual specification of the NDC boundary in particular domains – criminal justice, immigration, welfare, child protection – is left for the field to develop, though Section 3.3 develops the content moderation and media provenance case directly as an instance of aggregate NDC.

This claim has a well-known predecessor: Weizenbaum (1977) argued computers should never make certain decisions because they lack compassion and wisdom. The NDC's claim differs: it does not rest on what algorithmic systems lack in output quality (a question this article leaves unresolved), but on the accountability structure such decisions require, regardless of output quality.

The most consequential feature of the NDC is the accountability-capability paradox (Engin & Hand, 2025): as algorithmic systems become more capable, the pressure to delegate intensifies (Doshi-Velez et al., 2017). The better the system performs, the less any human authority can meaningfully second-guess it. Technical capability, or the perception of it, creates pressure to defer; and deference in governing decisions is delegation by another name. This pressure is already alive in current deployments of AI-assisted decision-making in bail determinations, immigration processing, welfare eligibility, and child protection referrals, where human oversight is increasingly nominal rather than substantive. Australia's Robodebt scheme illustrates the point: human review of automated debt notices was progressively removed, hundreds of thousands of debts were later found unlawful, and it ultimately took a Royal Commission to establish who was accountable (Royal Commission into the Robodebt Scheme, 2023). The NDC’s claim is that this pressure cannot be resolved by further technical improvement – only by institutional design that keeps human authority constitutively present, regardless of how capable the overseen systems have become. This position converges with AI safety research arguing that increasing capability intensifies rather than resolves the oversight problem: systems humans cannot meaningfully second-guess are precisely those over which meaningful human authority is hardest to maintain (Bengio et al., 2024).

Three objections must be addressed. The first is that human authorship of the frameworks within which algorithmic systems operate is sufficient: if humans design the rules, decisions taken within them are human decisions. For NDC decisions this fails, because those decisions are defined by the inadequacy of prior specification (Hart, 1961) – when principle is applied to unanticipated circumstances algorithmically, human authorship of the framework does not extend to the decision. The second objection is that an AI oversight layer restores the accountability the NDC requires. It does not: when AI System A reviews AI System B, the chain of authorisation is extended without being terminated (Engin & Hand, 2025) – adding complexity without adding human comprehension or responsibility. Recursive algorithmic oversight is precisely the institutional form the NDC rules out. The third objection is the strongest: well-designed algorithmic systems can be more consistent, easier to inspect, and easier to contest than the human bureaucracies they replace (Coglianese & Lehr, 2016; Kleinberg et al., 2018) – more answerable, in the terms Section 3.1 sets out. The NDC concedes this: consistency, inspectability, and contestability are genuine ability gains, and Section 2 requires them. But they fail for the same reason the second objection does – they improve what a system does, not who answers for it. The objection improves the answerability of outputs; the NDC concerns the accountability of an agent.

The four criteria above (fundamental rights, coercive power, irreversible consequences, contextual moral judgment) typically govern what can be called *decision-level NDC*: a single welfare determination, bail decision, or child protection referral. But the accumulation argument developed in Section 3.2 already established that legitimacy failures need not be located in any single decision; they can arise entirely from the aggregate pattern across many decisions, none of which individually meets a fundamental-rights or coercive-power threshold. Content moderation, algorithmic content ranking, and provenance determinations for AI-generated media are the clearest contemporary instances (Gillespie, 2018; Crick et al., 2025). No single moderation decision typically meets the decision-level NDC criteria, yet the cumulative pattern across millions of such decisions can narrow the epistemic commons democratic deliberation depends on – a systemic effect meeting the same underlying test the NDC applies at the individual level: no algorithmic system, and no individually contestable decision, can bear the accountability such a consequence requires. At its limit, the same mechanism – mass data collection, opaque inference, automated decisions with no human in the loop – becomes infrastructure for information control and unaccountable power (Barez et al., 2025). The NDC therefore operates at two levels. *Decision-level NDC* identifies individual decisions that cannot be delegated regardless of capability, requiring human authority constitutively present in the act of deciding. *Aggregate NDC* identifies patterns whose cumulative effect on collective epistemic or democratic capacity cannot be delegated either, requiring not presence in any single decision but continuous, population-level monitoring and correction, the hybrid institutional form Section 2.2 and Principle 5 already established for this reason (Suzor, 2019). Where *decision-level NDC* is violated by an absent decision-maker, *aggregate NDC* is violated by an absent monitor: an institution positioned to see, and empowered to correct, a pattern that no individual decision reveals.

The NDC applies on both axes. On the *over*-axis, it names the governing functions that private actors cannot legitimately exercise regardless of capability, because no other structure can bear the accountability they require. On the *with*-axis, it names the governing decisions states cannot legitimately delegate to algorithmic systems – human authority must be constitutively present, not as a formal residual.

**Principle 10 — The Non-Delegable Core**

*Certain governing decisions – determining fundamental rights, exercising coercive power, producing irreversible consequences, or requiring contextual moral judgment – cannot be legitimately delegated to algorithmic systems, regardless of technical capability. At the individual level, legitimacy requires a human agent present in the act of deciding, accountable for it and traceable to it. At the aggregate level, it requires continuous monitoring capable of catching what no single decision reveals.*

Together, these three conditions specify what human centricity requires in practice: authority that is accountable and traceable, held with judgment across time, and bounded by a Non-Delegable Core. Unlike Section 2's *ability* conditions, which specify what institutions must be able to do, these specify what governing arrangements must look like to those subject to them: visible enough to be understood, contestable enough to be challenged, and anchored enough in human authority to be genuinely accountable.

# 4. Conclusion: foundations for a field in formation

*Digital statecraft* was chosen as the organising concept for this field because each word carries necessary weight. 'Digital' names the full stack that must be governed – not merely its algorithmic surface but the data, connectivity, compute, and hardware supply chains through which political power is increasingly exercised. 'State' names the historically earned starting point: the institutional form that developed precisely the properties the algorithmic age now puts under pressure. And 'craft' names what governing it well has always demanded: to make adequate, not merely to arrange; to be legitimate, not merely functional: to be answerable, not merely automated.

Governing well in the algorithmic age requires two things simultaneously: *ability* as the precondition for the quality of public life, and *legitimacy* as the condition that distinguishes digital statecraft from authoritarian efficiency. Ability without legitimacy is capable oppression; legitimacy without ability is accountable impotence. The ten principles (Figure 1) are derived from this tension, each naming a condition whose absence produces its own identifiable, structural failure; diagnostic and severally correctable, not a single threshold that must be cleared in full. What is jointly necessary is not the ten holding at once but the two requirements themselves: a governance arrangement can satisfy *Systemic Coherence* while failing *Human Centricity and Civic Agency*, and that failure is the one principle's, not a verdict on the whole. Nor does every principle bear on every context – the *Non-Delegable Core*, for instance, has no purchase where no fundamental right, coercive power, or irreversible consequence is at stake, and where a condition is simply not material, its absence is an irrelevance rather than a failure. The systemic failure patterns discussed below sit at the coarser, requirement level: they follow from an entire requirement being neglected wholesale, not from any single principle's absence.

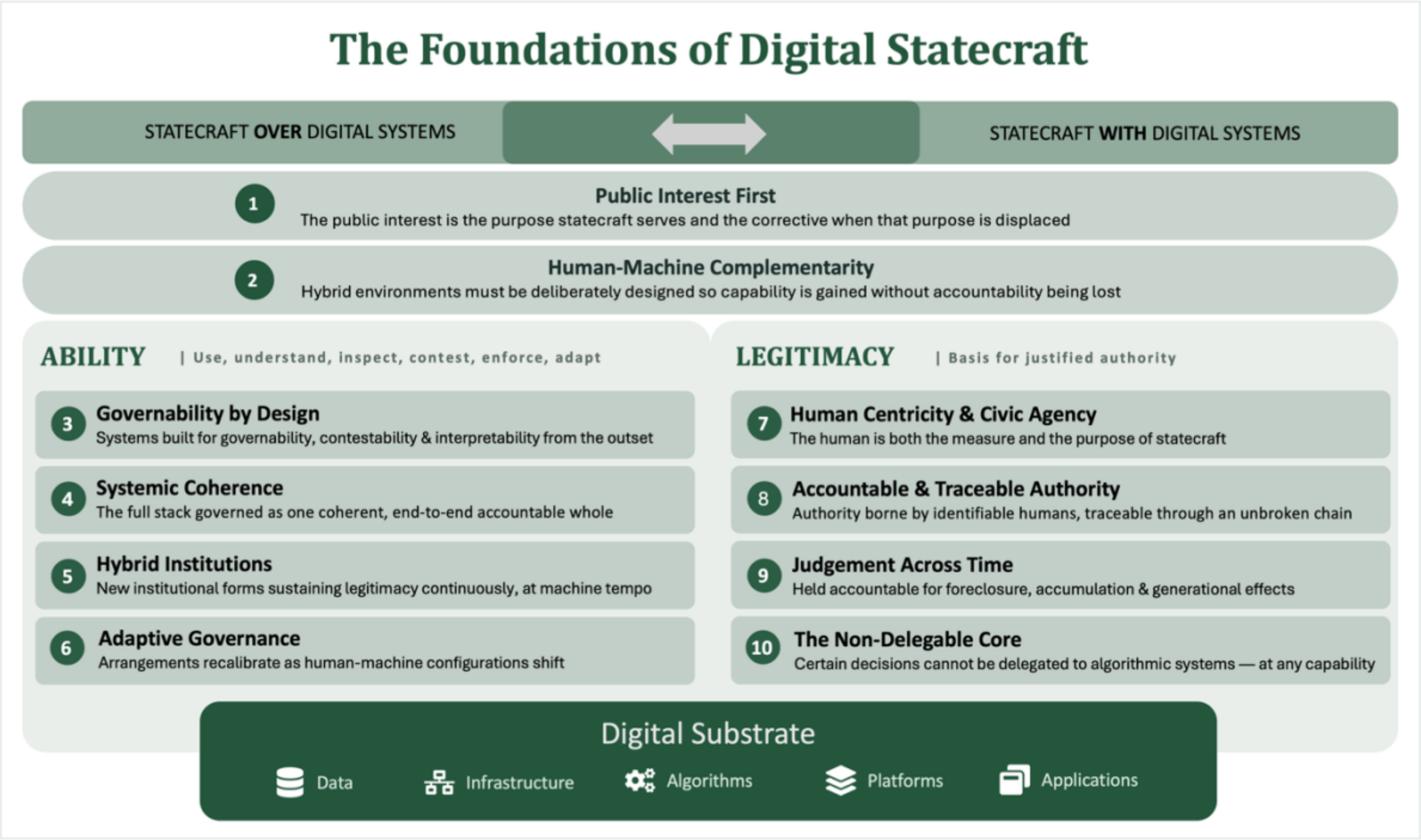

***Figure 1: The Foundations of Digital Statecraft – a summary***

These principles are distinctive in three respects. They are anchored in the governance problem rather than just the technical one: concerned with what institutions must be able to do, what governing arrangements must look like to those subject to them, and what system design must deliver to make both possible. Existing frameworks tend to treat transparency, fairness, and explainability as properties systems should exhibit; these principles treat them as conditions governing institutions must be able to exercise and citizens must be able to claim. They operate at the level of governance architecture rather than technical specification – concepts such as *Human-Machine Complementarity*, *Hybrid Institutions*, *Governability by Design*, and the *Non-Delegable Core* do not appear in existing frameworks, which remain largely addressed to the behaviour of systems rather than the structure of the authority governing them. And they are explicitly temporal: *Judgment Across Time* and *Adaptive Governance* treat legitimacy as a condition that must be sustained across shifting configurations and future publics, not merely satisfied at deployment.

The derivation offered here is foundational rather than exhaustive: a necessary basis for governing well in the algorithmic age, not a complete specification of everything it requires. Further work may identify additional requirements, refine these principles, or reveal tensions between them that the current level of abstraction does not surface.

The principles are deliberately stated at a level of generality that leaves their contextual specification open – and that openness constitutes the research agenda of the digital statecraft programme. Six lines of inquiry follow directly from the foundations established here: how the Non-Delegable Core boundary is identified in specific domains; what hybrid institutions look like in particular jurisdictions; how the *over*/*with* boundary is tracked as states acquire capability; what international cooperation architecture the traceability condition requires; how vertical traceability (as developed in Section 3.1) is secured in devolved, federal, and multi-level systems where digital and data policy competence is constitutionally distributed; and how accumulated effects of algorithmic governance are made visible and correctable over time. These are not admissions of incompleteness. They are what a foundational work is supposed to leave for the field it establishes, and they are offered here as the field's first commissions.

Where the conditions identified are not met, two failure modes follow. *Technocratic leviathan*, hyper-automated governance with centralised control and limited transparency, is the failure mode of ability deployed without legitimacy, already visible in automated welfare determinations, predictive policing, and algorithmic immigration processing where human oversight is nominal rather than substantive. *Platform sovereignty*, governing authority migrating to large technology companies beyond democratic reach, is the failure mode of state incapacity, carrying none of the accountability structures through which legitimate authority is exercised. Both are already present in current governance arrangements. The *digital statecraft* agenda exists to name them, specify their causes, and work toward governing arrangements that avoid them.

These foundational conditions derived are not equally accessible to all candidate governing actors – international bodies, states, sub-state and civil society actors, and private platform actors – and that asymmetry matters. Some deficits are contingent: absent because of insufficient investment or political will, but in principle remediable. Others are structural: features of an actor's institutional form that capability or commitment alone cannot resolve. The state is the only candidate for which the full set of legitimacy conditions is institutionally available: democratic mandate grounds traceable authority and consent, constitutional structures instantiate accountability and revisability, legal continuity sustains judgment across time. No other current actor possesses these properties in the same structural sense. The legitimacy disqualifications that platform actors face are structural: no amount of technical capability or voluntary commitment resolves the absence of democratic mandate or an accountability chain that terminates in shareholders rather than citizens. The state's deficits against the ability conditions are contingent – buildable, not inherent. But buildability is uneven: where states depend on infrastructure they do not control, face regulatory capacity that is outmatched by the actors they govern, or operate within international arrangements that concentrate standard-setting elsewhere, the pathways to building capacity are themselves constrained. For these states the ability conditions are contingent in principle but difficult in practice, a distinction that the *digital statecraft programme* must address if its foundations are to hold beyond the institutional contexts in which they were developed. Whether states can rebuild the *ability* and *legitimacy* conditions the algorithmic age requires – not recovering an industrial-era model of certainty, but constructing something adequate to conditions that model was never designed to face – is not a question this conceptual framework answers. It is the question it makes it possible to ask rigorously. And beneath it sits a deeper question that the programme holds open: whether the boundaries, scale, and affiliative basis of statehood itself are stable enough to reconstruct from, or whether the algorithmic age is already reconstituting them in ways that no single state can govern alone (Held, 1995; Zürn, 2018).

The destination – *resilient public futures* in which governing arrangements are networked, adaptive, and accountable, and technology serves human flourishing rather than displacing it – is not guaranteed. It is what the work is for. The ten principles are not that destination: they are the conditions its achievement requires. And the *craft* – the demanding, irreducibly human practice of holding *ability* and *legitimacy* together under conditions where both remain contested – is what *digital statecraft* demands of those who govern, and what this paper has sought to establish the foundations for.

**Acknowledgements:** This paper draws on an extended process of collaborative development from August 2025 to July 2026, including iterative writing sessions, wide-ranging conversations with The Digital Statecraft Academy fellows, faculty, experts, and guest contributors culminating in a week of intensive exchange during the Cambridge residential in April 2026. The ten principles of digital statecraft presented here develop an earlier formulation published in the DSA Founding Manifesto (Engin et al., 2025a). The views expressed in this document are those of the author(s) alone and do not represent the official position of any affiliated institution or office. We are grateful to members of the inaugural DSA Cambridge Fellowship cohort, beyond those who joined as co-authors, whose questions and challenges throughout the residential and beyond shaped this paper in ways the text alone cannot fully credit: Mohammed Alatoibi, Andrea Fernández Conde, Guy Gresham, Natalia Grincheva, Liesje Klomp, Bassam Al Mannai, Arturo Muente Kunigami, Jennifer Pougnet, Yannik Sassmann, Lillian Smith, Rosie Worth, Jazib Zahir. We are equally grateful to the following individuals for their part in the DSA's inaugural debates and conversations, which fed directly into the arguments presented here: Innar Liiv, Calum Handforth, Kennedy Mbeva, Felix Steffek, Miri Zilka, Sam Reynolds, Sadiq Jaffer, Sana Khareghani, Maria Luciana Axente, Rob Doubleday, Ray Eitel-Porter, Ade Famoti, Kenji Takeda, Abigail Sellen, Jennifer Hansen, Lina Hadburn, Julien Huppert, Inga Ulnicane, Alastair Beresford, Alice Hutchings, Ignacio Castro,

Nishanth Sastry, Anna Talas, Chang Liu, Tabitha Goldstaub, Jakob Mökander, Alastair Moore, Kate Jones, Hugo Leal, and Barbara Ubaldi. We are also grateful for discussions with Scott Hosking, Christopher Burr, and James Geddes at The Alan Turing Institute facilitated by Shane Connelly, Nicolas Guernion, and Matthew Corbett. Our particular thanks go to those who made the DSA's inaugural year possible: DSA Programme Manager Pinar Ozgen; Cambridge CSaP–DSA Fellowship Manager Christian Neubacher; and CSaP Director for Fellowships and Networks Nicky Buckley, for a full year of steady support in bringing the DSA together. Finally, we are hugely grateful to Liz Kendall, UK Secretary of State for Science, Innovation and Technology (2025-26), for her inspiring and encouraging remarks on the "importance and timeliness" of this initiative, and to Hetan Shah, Chief Executive of the British Academy, for his warm hospitality and welcome at the official launch of DSA in London. The list above cannot capture everyone who shaped this first year of the DSA. We remain deeply grateful to the much larger community of scholars, practitioners, and experts who lent their time, expertise, and encouragement to the Academy's inaugural year.

AI tools were used in the preparation of this manuscript in two ways, a use that itself sits inside the debate on AI-assisted scholarly production (Dwivedi et al., 2023). First, to surface connections across disciplinary literatures – a structural challenge that interdisciplinary scholarship routinely confronts but rarely acknowledges. All such connections were verified against primary sources and reviewed by human scholars from the relevant traditions. Second, to support the improvement of presentation, refining language and structure for a diverse interdisciplinary and cross-sector readership. The authors note that both uses reflect the *human-machine complementarity* the paper argues for – AI assistance expanding rather than replacing scholarly judgment, with human oversight and validation as the necessary counterpart. The authors reviewed and edited all AI-assisted text and take full responsibility for the content of the article.

**Data Availability Statement:** This is a conceptual paper and does not draw on primary or secondary empirical data.

**Funding Statement:** This research received no external funding.

# References

Acemoglu, D. & Johnson, S. (2023). *Power and progress: Our thousand-year struggle over technology and prosperity*. Hachette UK

Acharya, D. B., Kuppan, K., & Divya, B. (2025). Agentic AI: Autonomous intelligence for complex goals—A comprehensive survey. *IEEE Access, 13*, 18912-18936. https://doi.org/10.1109/ACCESS.2025.3532853

Anderson, B. (1983). Imagined communities: Reflections on the origin and spread of nationalism. Verso.

Aristotle. *Nicomachean Ethics, Book VI*.

Arrieta, A. B., Díaz-Rodríguez, N., Del Ser, J., Bennetot, A., Tabik, S., Barbado, A., ... & Herrera, F. (2020). Explainable Artificial Intelligence (XAI): Concepts, taxonomies, opportunities and challenges toward responsible AI. *Information Fusion, 58*, 82-115. https://doi.org/10.1016/j.inffus.2019.12.012

Barez, F., Friend, I., Reid, K., Krawczuk, I., Wang, V., Mökander, J., Torr, P., Morse, J., & Trager, R. (2025). Toward resisting AI-enabled authoritarianism. *Working paper, Oxford Martin AI Governance Initiative*, University of Oxford.

Barocas, S., Hardt, M., & Narayanan, A. (2023). *Fairness and machine learning: Limitations and opportunities*. MIT Press.

Beetham, D. (2013). *The legitimation of power*. Bloomsbury Publishing.

Bengio, Y., Hinton, G., Yao, A., Song, D., Abbeel, P., Darrell, T., ... & Mindermann, S. (2024). Managing extreme AI risks amid rapid progress. *Science, 384*(6698), 842-845. https://doi.org/10.1126/science.adn0117

Benites, R., Phipps, L., Watermeyer, R., & Crick, T. (2026). AI could transform research assessment — and some academics are worried. *Nature*. https://doi.org/10.1038/d41586-026-00124-8

Bertelli, A. M. and Peter, J. (2013), Statecraft and the Art of Governing, Public Policy Investment: Priority-Setting and Conditional Representation In: *British Statecraft*. Oxford University Press. https://doi.org/10.1093/acprof:oso/9780199663972.003.0003

Bovens, M. (2007). Analysing and assessing accountability: A conceptual framework. *European Law Journal, 13*(4), 447-468. https://doi.org/10.1111/j.1468-0386.2007.00378.x

Bratton, B. H. (2026). *The Stack: On Software and Sovereignty*. MIT Press.

Brownsword, R. (2008). *Rights, regulation, and the technological revolution*. Oxford University Press.

Caldwell, L. K. (1996). The state as a work of art: Statecraft for the 21st century. *PS: Political Science & Politics, 29*(4), 657-664. https://doi.org/10.2307/420788

Calo, R. (2017). Artificial intelligence policy: a primer and roadmap. *UC Davis Law Review, 51*, 399. https://digitalcommons.law.uw.edu/faculty-articles/640

Cavoukian, A. (2009). Privacy by design: The 7 foundational principles. Information & Privacy Commissioner of Ontario, Canada.

Charalabidis, Y., Flak, L. S., & Pereira, G. V. (2022). Scientific foundations of digital governance and transformation. In: *Public Administration and Information Technology*. Springer.

Citron, D. K. (2007). Technological due process. *Washington University Law Review, 85*, 1249. https://openscholarship.wustl.edu/law_lawreview/vol85/iss6/2

Coeckelbergh, M. (2023). Democracy, epistemic agency, and AI: political epistemology in times of artificial intelligence. *AI and Ethics, 3*(4), 1341-1350. https://doi.org/10.1007/s43681-022-00239-4

Coglianese, C., & Lehr, D. (2016). Regulating by robot: Administrative decision making in the machine-learning era. *The Georgetown Law Journal. LJ*, 105, 1147.

Cohen, J. E. (2019). *Between truth and power*. Oxford University Press.

Collingridge, D. (1980). *The social control of technology*. Frances Pinter.

Crick, T., Chadwick, A., Strong, C., Jiga-Boy, G., Madsen, J., Burkhardt, J., Roozenbeek, J., Harris, L., Horvath, L., et al. (2025). Deepfakes and media literacy. Government Office for Science, UK Government. https://www.gov.uk/government/publications/deepfakes-and-media-literacy/deepfakes-and-media-literacy

Daníelsson, J. (2002). The emperor has no clothes: Limits to risk modelling. *Journal of Banking & Finance, 26*(7), 1273-1296. https://doi.org/10.1016/S0378-4266(02)00263-7

Diakopoulos, N. (2016). Accountability in algorithmic decision making. *Communications of the ACM, 59*(2), 56-62. https://doi.org/10.1145/2844110

Doshi-Velez, F., Kortz, M., Budish, R., Bavitz, C., Gershman, S., O'Brien, D., ... & Wood, A. (2017). Accountability of AI under the law: The role of explanation. arXiv preprint https://doi.org/10.48550/arXiv.1711.01134

Dunleavy, P., & Margetts, H. (2025). Data science, artificial intelligence and the third wave of digital era governance. *Public Policy and Administration, 40*(2), 185-214. https://doi.org/10.1177/09520767231198737

Dwivedi, Y. K., et al. (2021). Artificial Intelligence (AI): Multidisciplinary Perspectives on Emerging Challenges, Opportunities, and Agenda for Research, Practice and Policy. *International Journal of Information Management, 53*, 101994. https://doi.org/10.1016/j.ijinfomgt.2019.08.002

Dwivedi, Y. K., et al. (2023). "So what if ChatGPT wrote it?" Multidisciplinary perspectives on opportunities, challenges and implications of generative conversational AI for research, practice and policy. *International Journal of Information Management, 71*, 102642. https://doi.org/10.1016/j.ijinfomgt.2023.102642

Engin, Z. (2026). Human-AI Governance (HAIG): A trust-utility approach. *Journal of Responsible Technology, 26*, 100167. https://doi.org/10.1016/j.jrt.2026.100167

Engin, Z., Crowcroft, J., Hand, D., & Treleaven, P. (2025b). The Algorithmic State Architecture (ASA): an integrated framework for AI-enabled government. arXiv preprint. https://doi.org/10.48550/arXiv.2503.08725

Engin, Z., & Hand, D. (2025). The Non-Delegable Core: Designing Legitimate Oversight for Agentic AI. Zenodo. https://doi.org/10.5281/zenodo.16685275

Engin, Z., & Hand, D. (2026). Towards adaptive categories: Dimensional governance for agentic AI. *Data & Policy*, *8*, e38. doi:10.1017/dap.2026.10083

Engin, Z., Hand, D., Verhulst, S., Gardner, E., Crowcroft, J., Treleaven, P., & Gordon, T. (2025a). The Digital Statecraft Manifesto v1.0: The art, science, and practice of governance in the algorithmic age. The Digital Statecraft Academy (DSA). https://digitalstatecraft.academy/the_digital_statecraft_manifesto_v1-0/
Engin, Z., & Treleaven, P. (2019). Algorithmic government: Automating public services and supporting civil servants in using data science technologies. *The Computer Journal, 62*(3), 448-460. https://doi.org/10.1093/comjnl/bxy082
Eubanks, V. (2018). *Automating inequality: How high-tech tools profile, police, and punish the poor.* St Martin's Press.
Floridi, L., Cowls, J., Beltrametti, M., Chatila, R., Chazerand, P., Dignum, V., ... & Vayena, E. (2018). AI4People—An ethical framework for a good AI society: Opportunities, risks, principles, and recommendations. *Minds & Machines, 28*(4), 689-707. https://doi.org/10.1007/s11023-018-9482-5
Fourcade, M., & Gordon, J. (2020). Learning like a state: Statecraft in the digital age. *Journal of Law and Political Economy, 1*(1), 78. https://doi.org/10.5070/LP61150258
Gillespie, T. (2018). *Custodians of the Internet: Platforms, Content Moderation, and the Hidden Decisions That Shape Social Media*. Yale University Press.
Grimmelikhuijsen, S., & Meijer, A. (2022). Legitimacy of algorithmic decision-making: Six threats and the need for a calibrated institutional response. *Perspectives on Public Management and Governance, 5*(3), 232-242. https://doi.org/10.1093/ppmgov/gvac008
Gohwong, S. (2026). Chinese net states and platform-based governance in China: Implications for comparative net-state analysis. *Procedia of Multidisciplinary Research*, 4(2).
Habermas, J. (2015). Between facts and norms: Contributions to a discourse theory of law and democracy. John Wiley & Sons.
Hand, D. J. (2026). *What's the Question?: Deciding What You Really Want to Know*. CRC Press.
Hart, H. L. A. (1961). *The Concept of Law*. Oxford University Press.
Haug, N., Dan, S., & Mergel, I. (2024). Digitally-induced change in the public sector: a systematic review and research agenda. *Public Management Review, 26*(7), 1963-1987. https://doi.org/10.1080/14719037.2023.2234917
Held, D. (1995). *Democracy and the global order: From the modern state to cosmopolitan governance*. Stanford University Press.
Hildebrandt, M. (2015). Smart technologies and the end (s) of law: novel entanglements of law and technology. In: *Smart Technologies and the End(s) of Law*. Edward Elgar Publishing.
Hildebrandt, M. (2020). *Law for computer scientists and other folk*. Oxford University Press.
Hughes, L., Dwivedi, Y. K., Malik, T., Shawosh, M., Albashrawi, M. A., Jeon, I., Dutot, V., Appanderanda, M., Crick, T., De, R., Fenwick, M., Senali, M., Jurcys, P., Kar, A., Li, K., Mutasa, S., Samothrakis, S., Walton, P., & Wade, M. (2025). AI Agents and Agentic Systems: A Multi-Expert Analysis. *Journal of Computer Information Systems, 65*(4), 489-517. https://doi.org/10.1080/08874417.2025.2483832
ISO/IEC 42001. (2023). Information technology – Artificial intelligence – Management system. International Organization for Standardization.
Janssen, M., Hartog, M., Matheus, R., Yi Ding, A., & Kuk, G. (2022). Will algorithms blind people? The effect of explainable AI and decision-makers' experience on AI-supported decision-making in government. *Social Science Computer Review, 40*(2), 478-493. https://doi.org/10.1177/0894439320980118
Jonas, H. (1984). *The imperative of responsibility: In search of an ethics for the technological age*. University of Chicago Press.
Kalema, N. L. (2024). The 'Digital Transformation for Development' anti-politics machine: A case study on global digital development governance and public-sector digital transformation in Uganda. *Policy & Internet, 16*(4), 750-763. https://doi.org/10.1002/poi3.436
Kleinberg, J., Ludwig, J., Mullainathan, S., & Sunstein, C. R. (2018). Discrimination in the Age of Algorithms. *Journal of legal analysis*, 10, 113-174. https://doi.org/10.1093/jla/laz001
Koshiyama, A., & Engin, Z. (2019). Algorithmic impact assessment: Fairness, robustness and explainability in automated decision-making. UCL presentation.
Leveson, N. G. (2016). *Engineering a safer world: Systems thinking applied to safety*. MIT Press.
Lipton, Z. C. (2018). The mythos of model interpretability. *Communications of the ACM, 61*(10), 36-43. https://doi.org/10.1145/3233231
Lnenicka, M., Clarinval, A., Nikiforova, A., Rudmark, D., Luterek, M., Symeonidis, D. and Bolívar, M.P.R., (2026). Artificial intelligence in policymaking: Mapping integration gaps across the public policy cycle. *Government Information Quarterly, 43*(2), 102138. https://doi.org/10.1016/j.giq.2026.102138

Lu, J., Liu, A., Dong, F., Gu, F., Gama, J., & Zhang, G. (2018). Learning under concept drift: A review. *IEEE Transactions on Knowledge and Data Engineering, 31*(12), 2346-2363. https://doi.org/10.1109/TKDE.2018.2876857
Luna-Reyes, L. F., & Gil-Garcia, J. R. (2014). Digital government transformation and internet portals: The co-evolution of technology, organizations, and institutions. *Government Information Quarterly, 31*(4), 545-555. https://doi.org/10.1016/j.giq.2014.08.001
Mann, M. (1984). The autonomous power of the state: its origins, mechanisms and results. *European Journal of Sociology/Archives Européennes de Sociologie, 25*(2), 185-213. https://doi.org/10.1017/S0003975600004239
Mazzucato, M. (2021). *Mission economy: A moonshot guide to changing capitalism*. Penguin UK.
Meadows, D. (2008). *Thinking in systems*. Chelsea Green Publishing.
METI. (2021). Governance innovation ver. 2: A guide to designing and implementing agile governance. Ministry of Economy, Trade and Industry, Government of Japan.
Miller, D. (1995). *On nationality*. Oxford University Press.
Mittelstadt, B. D., Allo, P., Taddeo, M., Wachter, S., & Floridi, L. (2016). The ethics of algorithms: Mapping the debate. *Big Data & Society, 3*(2). https://doi.org/10.1177/2053951716679679
Moon, D. S., & Evans, T. (2017). Welsh devolution and the problem of legislative competence. *British Politics, 12*(3), 335–360. https://doi.org/10.1057/s41293-016-0043-3
Moore, M. H. (2013). *Recognizing public value*. Harvard University Press.
Mökander, J., Schuett, J., Kirk, H. R., & Floridi, L. (2024). Auditing large language models: a three-layered approach. AI and Ethics, 4(4), 1085–1115. https://doi.org/10.1007/s43681-023-00289-2
Mulgan, G. (2017). *Big mind: How collective intelligence can change our world*. Princeton University Press.
Mulligan, D. K., & Bamberger, K. A. (2019). Procurement as policy: Administrative process for machine learning. *Berkeley Technology Law Journal, 34*. https://doi.org/10.2139/ssrn.3464203
Nikiforova, A., Rodrigues, M. P., & Tangi, L. (2026). Proactive Public Services in the EU: Readiness, maturity, and pathways to implementation. *Publications Office of the European Union, Luxembourg.* https://data.europa.eu/doi/10.2760/4554259, JRC146641.
NIST AI RMF – National Institute of Standards and Technology. (2023). Artificial intelligence risk management framework (AI RMF 1.0). U.S. Department of Commerce.
Noveck, B. S. (2015). *Smart citizens, smarter state: The technologies of expertise and the future of governing*. Harvard University Press.
OECD. (2021). Recommendation of the Council for agile regulatory governance to harness innovation (OECD/LEGAL/0464). OECD.
OED (n.d.). Statecraft (n.). In: *Oxford English Dictionary*. Retrieved July 12, 2026, from https://www.oed.com/dictionary/statecraft_n
O'Neil, C. (2016). *Weapons of Math Destruction: How Big Data Increases Inequality and Threatens Democracy*. Crown.
Parnas, D. L. (1972). On the criteria to be used in decomposing systems into modules. *Communications of the ACM, 15*(12), 1053-1058. https://doi.org/10.1145/361598.361623
Pasquale, F. (2015). *The black box society: The secret algorithms that control money and information*.. Harvard University Press.
Pasquale, F. (2020). *New laws of robotics*. Harvard University Press.
Pettit, P. (1997). *Republicanism: A theory of freedom and government*. Oxford University Press.
Plantin, J. C., Lagoze, C., Edwards, P. N., & Sandvig, C. (2018). Infrastructure studies meet platform studies in the age of Google and Facebook. *New Media & Society, 20*(1), 293-310. https://doi.org/10.1177/1461444816661553
Pohle, J., & Thiel, T. (2020). Digital sovereignty. *Internet Policy Review, 9*(4). https://doi.org/10.14763/2020.4.1532
Power, M. (2007). *Organized uncertainty: Designing a world of risk management*. Oxford University Press.
Rahman, K. S. (2018). Regulating informational infrastructure: Internet platforms as the new public utilities. *Georgetown Law and Technology Review, 2*(2). https://ssrn.com/abstract=3220737
Raji, I. D., Smart, A., White, R. N., Mitchell, M., Gebru, T., Hutchinson, B., ... & Barnes, P. (2020, January). Closing the AI accountability gap: Defining an end-to-end framework for internal algorithmic auditing. In: *Proceedings of the 2020 Conference on Fairness, Accountability, and Transparency* (pp. 33-44). https://doi.org/10.1145/3351095.3372873
Rawls, J. (1971). *A theory of justice*. Harvard University Press.

Renan, E. (1882). Qu'est-ce qu'une nation? [What is a nation?]. Fordham Internet History Sourcebooks. https://sourcebooks.web.fordham.edu/mod/1882Renan-nation-eng.asp
Royal Commission into the Robodebt Scheme (2023). Report of the Royal Commission into the Robodebt Scheme. *Commonwealth of Australia*. https://robodebt.royalcommission.gov.au/publications/report
Rouvroy, A., & Berns, T. (2013). Algorithmic governmentality and prospects of emancipation: Disparateness as a precondition for individuation through relationships?. *Réseaux, 177*(1), 163-196. https://doi.org/10.3917/res.177.0163
Scottish Parliament Information Centre (SPICe). (2024). Artificial Intelligence and Healthcare in Scotland (SB 24-34). Scottish Parliament. https://www.parliament.scot/chamber-and-committees/research-prepared-for-parliament/research-briefings/2024/6/25/sb-2434
Selbst, A. D., Boyd, D., Friedler, S. A., Venkatasubramanian, S., & Vertesi, J. (2019). Fairness and abstraction in sociotechnical systems. In: *Proceedings of the 2019 Conference on Fairness, Accountability, and Transparency* (pp. 59-68). https://doi.org/10.1145/3287560.3287598
Senedd Research. (2024). Generative AI – an overview (Research Briefing 24-14). Welsh Parliament. https://research.senedd.wales/media/s4ldqlwj/24-14-generative-ai-en.pdf
Skorpen, G. S., & Reinertsen, H. (2025). Making space for digital statecraft: The work of consultancy models in an audit of police digitalisation. *Nordic Journal of Science and Technology Studies, 13*(1). https://doi.org/10.5324/njsts.v13i1.5858
Spruyt, H. (1994). *The sovereign state and its competitors: An analysis of systems change*. Princeton University Press.
Suleyman, M. (2023). *The coming wave: AI, power, and our future*. Random House.
Suzor, N. P. (2019). *Lawless: The Secret Rules That Govern Our Digital Lives*. Cambridge University Press.
Tilly, C. (1992). *Coercion, capital, and European states, AD 990-1992*. Oxford University Press.
Van Dijck, J., Poell, T., & De Waal, M. (2018). *The platform society: Public values in a connective world*. Oxford University Press.
van Kranenburg, R. (2024). *Statecraft and Policymaking in the Age of Digital Twins*. Springer.
Wachter, S., Mittelstadt, B., & Russell, C. (2017). Counterfactual explanations without opening the black box: Automated decisions and the GDPR. *Harvard Journal of Law & Technology, 31*, 841-887. https://doi.org/10.2139/ssrn.3063289
Watermeyer, R., Phipps, L., Benites, R., & Crick, T. (2025). REF-AI: Exploring the potential of generative AI for REF2029. Centre for Higher Education Transformations, University of Bristol and Jisc. https://chet.bristol.ac.uk/research/ref-ai-project/
Watson, D. S., Mökander, J., & Floridi, L. (2024). Competing narratives in AI ethics: a defense of sociotechnical pragmatism. *AI & Society*. https://doi.org/10.1007/s00146-024-02128-2
Weber, M. (2019). *Economy and society: A new translation*. Harvard University Press.
Weizenbaum, J. (1977). *Computer power and human reason: From judgment to calculation*. W. H. Freeman.
WEF (2018). Agile governance: Reimagining policy-making in the Fourth Industrial Revolution. World Economic Forum.
Widmer, G., & Kubat, M. (1996). Learning in the presence of concept drift and hidden contexts. *Machine Learning, 23*(1), 69-101. https://doi.org/10.1007/BF00116900
Yeung, K., & Lodge, M. (Eds.). (2019). *Algorithmic regulation*. Oxford University Press.
Zarsky, T. (2016). The trouble with algorithmic decisions: An analytic road map to examine efficiency and fairness in automated and opaque decision making. *Science, Technology, & Human Values, 41*(1), 118-132. https://doi.org/10.1177/0162243915605575
Zuboff, S. (2023). The age of surveillance capitalism. In: *Social Theory Re-wired* (pp. 203-213). Routledge.
Zürn, M. (2018). *A theory of global governance: Authority, legitimacy, and contestation*. Oxford University Press.